\documentclass{elsarticle}

\usepackage{amssymb}

\usepackage[figuresright]{rotating}

\makeatletter
\newlength \figwidth
\if@twocolumn
\else
\fi
\makeatother

\begin{document}

\begin{frontmatter}




\title{Modeling Heterogeneity in Networks \\
using Uncertainty Quantification Tools}


\author[a1]{Karthikeyan Rajendran}
\author[a1]{Andreas C. Tsoumanis}
\author[a2]{Constantinos I. Siettos}
\author[a3]{Carlo R. Laing}
\author[a1,a4]{Ioannis G. Kevrekidis}

\address[a1] {Department of Chemical and Biological Engineering, Princeton University, Princeton, NJ, USA}
\address[a2] {School of Applied Mathematics and Physical Sciences, NTUA, Athens, Greece}
\address[a3] {Institute for Natural and Mathematical Sciences, Massey University, Auckland, New Zealand}
\address[a4] {Program in Applied and Computational Mathematics, Princeton University, Princeton, NJ, USA}

\begin{abstract}
Using the dynamics of information propagation on a network as our illustrative example,
we present and discuss a systematic approach to quantifying {\em heterogeneity} and its
propagation that borrows established tools from Uncertainty Quantification.
The crucial assumption underlying this mathematical and computational
``technology transfer" is that the evolving states of the nodes in a
network quickly become correlated with the corresponding node ``identities":
features of the nodes imparted by the network structure (e.g. the node degree, the node
clustering coefficient).
The node dynamics thus depend on {\em heterogeneous} (rather than {\em uncertain})
parameters, whose distribution over the network results from the network structure.
Knowing these distributions
allows us to obtain an efficient coarse-grained representation of the network state
in terms of the expansion coefficients in suitable orthogonal polynomials.
This representation is closely related to mathematical/computational tools
for uncertainty quantification (the Polynomial Chaos approach and its
associated numerical techniques).
The Polynomial Chaos coefficients provide a set of
good collective variables for the observation of dynamics on a network,
and subsequently, for the implementation of reduced dynamic models of it.
We demonstrate this idea by performing coarse-grained computations of the
nonlinear dynamics of
information propagation on our illustrative network model
using the Equation-Free approach \cite{Kevr04Equation-free:}.
\end{abstract}


\end{frontmatter}


\section{Introduction}
\label{sec:intro}

Our purpose in this paper is to establish a link between Uncertainty Quantification
in dynamical systems depending on (uncertain) parameters, and Heterogeneity Quantification
in dynamical systems consisting of (many) individual dynamical units coupled in a network.
These units differ in the way they are structurally linked to form the network; their
``structural identities" are heterogeneous, and the distribution of
these heterogeneities derives from the network structure itself.

Network models are increasingly being used to study large, complex real life systems
in a variety of contexts such as the internet,
chemical and biochemical reaction networks, social networks and more
\cite{Falo99power-law,Newm02random,Aren06synchronizationa,
Binz09topology,Lain09dynamics}.
A network is a mathematical representation of individual subsystems called nodes (or vertices),
which are connected to one another through edges (or links).
In the specific example of a social network the nodes represent people, while the edges connecting
them represent relationships (friendships, coauthorships, etc.) between them.
The following references provide useful reviews of basic network concepts and of
the study of dynamics of evolving networks
\cite{Albe02statistical,Doro02evolution,Newm03structure,Bocc06complex,Newm06structure,Barr08dynamical}.
Dynamic evolution in a network context can be broadly classified into two categories: dynamics {\em of } networks,
and dynamics {\em on} networks.
The former refers to problems where
the network structure itself changes over time according to some pre-specified rules of
evolution; the latter refers to problems where there is a static network structure and the states of the nodes
(variables associated with the nodes) evolve following pre-specified rules.
These two categories are not mutually exclusive: one
can model systems where both the nodal states and the network structure change over time
(the term ``adaptive networks" has been used for this combination, \cite{Gros08robust}).

In this work, we are interested in studying dynamics {\em on} networks
at a coarse-grained level using a systematic model reduction framework
called the {\em Equation-Free} approach \cite{Kevr04Equation-free:,Kevr03Equation-free}.
In this approach, short bursts of simulation at the (``fine") level
of nodes and edges using the detailed rules of dynamic evolution of the problem
are performed
in order to estimate enough information to carry out computational tasks at
a more coarse-grained level.
The Equation-Free approach has been, in the past, successfully implemented for a variety of
specific network models \cite{Raje11coarse,Tsou12coarse-graining,Bold12Equation-Free}.
The success of this approach rests heavily on (a) defining a suitable set of
coarse observables in terms of which a closed, reduced description of the
evolution on the network may theoretically be obtained,
and (b) the ability to convert back and forth between the
two levels of description of the system - the ``fine" and the ``coarse-grained" levels.
Thus, one of the most important steps in coarse graining using this approach
is the selection of appropriate coarse variables.

We propose a useful (and hopefully efficient) representation
of coarse variables specifically for models describing dynamics {\em on} networks, when
the states of the nodes of the network {\em quickly} become
correlated with features of the nodes imparted by the network, such as the node degree
and/or the node clustering coefficient.
In such cases we show that one can expand the function representing the dependence
of node state to node structural identity in
terms of suitable orthogonal polynomials depending on the distribution
from which the node feature(s) is sampled (i.e. depending on the topology of the
underlying network).
This idea is analogous to the  study of the  effects of random parameters
with a known distribution on uncertain dynamical systems \cite{Ghan91stochastic},
and we will discuss this analogy in more detail below.

In order to illustrate these ideas we consider a simple agent-based
model of opinion propagation where the agents are connected by a social network;
simulations of this model indicate that the states of the agents
become quickly correlated to the connectivity degrees of these agents as nodes in the network.
The paper is organized as follows:
Our illustrative model is described in Sec.~\ref{sec:model}, along with
a quick overview of its nonlinear dynamic behavior.
Sec.~\ref{sec:cr} defines and describes the coarse representation
that forms the basis of our computational reduced model.
A few details of the coarse variable
description are relegated to the Appendix, in order to maintain the
simplicity and flow of the discussion.
A brief outline of the Equation-Free approach employed to
computationally implement the reduced model is described in
Sec.~\ref{sec:ef}, along with the results of Equation-Free computations.
In Sec.~\ref{sec:conc}, we summarize the results and briefly discuss possible
extensions and directions for future work.

\section{The Model}
\label{sec:model}

We consider a simple, illustrative model of information propagation
in a population of agents connected by a social network structure,
as described in \cite{Tsou10Equation-Free}.
We briefly describe the individual-based rules of evolution of the model,
that is, the dynamics at the level of agents/nodes and connections
between them.
We consider a population of $N$ agents (numbered from $1$ to $N$)
connected in a network $G$ with corresponding adjacency matrix $A$
(i.e., if agents $i$ and $j$ are connected to each other, $A(i,j)$ is
$1$; if not, $A(i,j)=0$).
Each agent $i$ is associated with a single scalar variable $X_i$, that
denotes the emotional state of the agent.
The variables $X_i$ are bounded between $-1$ and $1$.
The agents receive
``public information" at discrete time instances from the external
environment; the information arrival times are modeled as
a Poisson process.
In addition to this public information, the agents also receive ``private
information" from their social environment (from their immediate neighbors in the
social network).
The emotional state of each agent changes according to the following
rules:
\begin{enumerate}

  \item The emotional state of each agent decays exponentially in time to
  zero. Thus, if $\Delta t$ is the time interval between two consecutive
  arrivals of information, the emotional state of every agent decreases
  by a factor of $exp(- \gamma \Delta t)$ in this time interval,
  where $\gamma$ is the decay parameter.

  \item Public information is classified as ``good" or ``bad",  both
  modeled as Poisson processes with arrival rates $\nu^+$ and $\nu^-$ respectively.
  Whenever agents receive good (respectively, bad) public information,
  their emotional state jumps by a small, finite positive constant (resp., negative)
  value $\epsilon^+$ (resp., $\epsilon^-$).

  \item The occurrence of private information transfer through the social network
  is also  modeled as a Poisson process, similar to public information arrivals;
  but private information arrival rates are taken to be proportional to the connectivity degree of the receiving agent. Hence, the time interval between arrivals of private information
  for agent $i$ with degree $d_i$ is modeled as a Poisson process
  with mean $\alpha d_i$ where $\alpha$ is a positive constant.
  Whenever an agent is due to receive private information he/she compares his/her state with the state of an immediate neighbor, chosen at random. If the emotional state of this
  neighbor is larger (resp., smaller) than the emotional state of the agent, then a constant
  parameter $e^+$ (resp., $e^-$) is added to the agent's emotional state. Note that $e^+$ and $e^-$ are positive and negative respectively.

  \item In order to ensure that the emotional states stay within the lower and
  upper bounds of $-1$ and $1$, whenever the emotional state of an agent reaches
  these bounds, the state is locked to that value for the remaining fragment of the time interval,
  regardless of the arrival of new information.
  This can be thought of as a saturation event where the agent finalizes his/her decision.

\end{enumerate}

\begin{figure*}
\begin{center}
\resizebox{0.72\figwidth}{!}{%
  \includegraphics{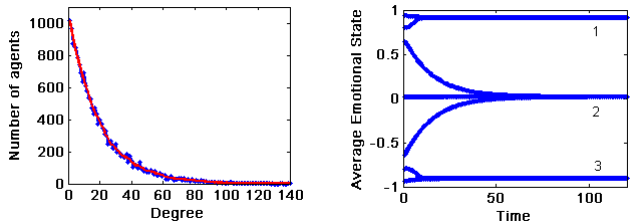}
}
\caption{
\textbf{Left:} The degree histogram of the illustrative network structure
(i.e., the number of agents with a given degree plotted versus the degree)
used in numerical simulations is plotted as (blue) dots.
The theoretical degree histogram from which our particular
degree sequence is sampled is plotted as a (red) solid line.
\textbf{Right:}
Evolution in time of the average emotional state of all the agents
in the network for the parameter settings listed in the text.
Depending on the initial conditions, the system reaches one of three stable steady states
(marked $1$, $2$ and $3$).
}
\label{fig:Eav}
\end{center}
\end{figure*}

We used $N=20,000$ agents in the model, and the network was constructed so that
the degree distribution is a modified form of the discrete geometric distribution
with degrees up to an upper limit of $140$.
The parameter $p$ of this truncated geometric degree
distribution is taken to be $0.05$.
The degree histogram for the network structure that we use for numerical simulations
is plotted as (blue) dots in the left plot of Fig.~\ref{fig:Eav}.
The theoretical histogram corresponding to the truncated degree distribution is
plotted as a (red) curve for comparison.
The following values were used for the remaining parameters of the model
for the purpose of numerical simulations: $\nu^+ = 20; \nu^- = 20; \epsilon^+ = 0.075;
\epsilon^- = -0.072; e^+ = 0.033; e^- = -0.035; \alpha = 2; \gamma = 0.5$.

\subsection{Nonlinear model behavior}
\label{ss:md}

The dynamical behavior of the model is described in considerable detail in
\cite{Tsou10Equation-Free}.
We briefly recount some basic features of these dynamics here.
Direct simulations, using the model rules, initialized at
different initial conditions are presented.
The state of the system at any moment in time is completely specified
by the states of each of the $20,000$ agents in the network.
Certain properties of the system can be best conveyed through chosen
collective observables that are hopefully representative of the
overall dynamics of the system.
The average emotional state of all network agents is one such
observable of interest.
The evolution of average emotional state of all the agents in the
network from various distinct initializations is shown in Fig.~\ref{fig:Eav};
for simplicity, the states
of all the agents were initialized uniformly at fixed values over the network (but different
fixed values for each initialization).
The figure shows that the system reaches one of three stable
steady states depending on the initial conditions;
parameter settings leading to a single stable, or to two stable stationary states
also exist in a detailed bifurcation diagram and have been discussed in \cite{Tsou10Equation-Free}.

\begin{figure*}
\begin{center}
\resizebox{0.72\figwidth}{!}{%
  \includegraphics{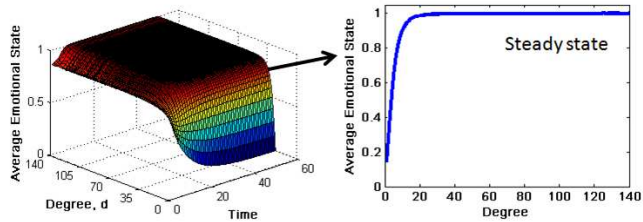}
}
\caption{
A $3D$ plot of the evolution of the average emotional state of agents with a given degree versus the
degree is shown evolving over time on the left.
The steady state of this average emotional state versus degree
is shown on the right.
The initial condition for the simulation gave all agents
a uniform emotional state of $0.8$.
}
\label{fig:Ecorr}
\end{center}
\end{figure*}

\section{Coarse representation}
\label{sec:cr}

To obtain a reduced model, one must first select
a set of coarse variables that accurately
capture the long-term evolution of the system.
To motivate our choice, we examined
the detailed profiles of system states along an ensemble of trajectories like the ones summarized in
Fig.~\ref{fig:Eav}.
For this model, we observed that the states of the  agents {\em quickly become
highly correlated} with their degrees.
Fig.~\ref{fig:Ecorr} shows the evolution in time of the average state
of all agents in a degree class (i.e., all agents having the same degree)
as a function of the degree.
The curve evolves smoothly in time, and it was demonstrated in \cite{Tsou10Equation-Free}
that such a correlation could indeed be used to obtain a reduced description of the model.
A method of ``binning" was employed in that work to construct good collective variables:
this involved partitioning the network nodes into different groups (based on the node degrees)
and required $80$ such groups, leading to 80 coarse variables.

We now realize that what, in that paper, was an {\em ad hoc} reduction is just a special
case of a very general, and potentially powerful approach to systematically reducing the
dynamics of large/complex networks.
Instead of following the behavior of each agent,
we \emph{exploit} the (assumed) fact that structurally similar agents have
similar behavior and can be tracked together.
We assume here that the key variable that describes structural similarity
is \emph{the degree} of the agent in the network.
The extension to more identities encompassing more structural features,
and to possibly {\em intrinsic} heterogeneities between the agents, over and
beyond the network-imparted {\em structural} heterogeneities will be
addressed in our Discussion section.
Finding the relationship between agent structural characteristics (here, agent degrees)
and agent states generates a coarse description,
whereby the system state can be encoded in (hopefully drastically) fewer independent variables.

Let $x_d$ denote the average emotional state of all agents with degree
$d$, where $d \in [1,140]$.
Then, the curve in Fig.~\ref{fig:Ecorr} can be represented by
a function $x_d = f(d)$.
Note that since the degrees of the nodes in the network range between $1$ and $140$,
this ``curve" is, in effect, a vector of $140$ values.
An intuitive approach to obtaining a reduced description of the function $f$ is to
expand it in terms of suitable basis functions.
Consider, for example, some set of orthogonal polynomials $p_i$ as basis
functions, where $i \in [1,k]$.
One can then expand the function $f$ in terms of these basis functions as
\begin{equation}
f(d)=\sum_{i=1}^{k} c_i p_i(d).
\label{eq:B}
\end{equation}

If the $k$ basis functions suffice to accurately capture the shape of the function,
one can then use the $k$ coefficients $c_i$ as the coarse representation of $f(d)$.
Once the polynomials are selected, the common method to evaluate the coefficients is to
find the optimal values of these coefficients that minimize the residual defined below:

\begin{equation}
<f(d)-\sum_{i=1}^{k} c_i p_i(d), f(d)-\sum_{i=1}^{k} c_i p_i(d)>_{w(d)}.
\label{eq:MR}
\end{equation}
Note that the inner product in the residual is defined with respect to a weight
function $w(d)$.
In our example, the values of the function $f(d)$ which describes
the average state for every degree class is computed {\em using different
numbers of agents at each $d$}.
It is therefore reasonable to use a weight function which mirrors this
sampling.
In simple terms, if there are many agents with a given degree $d$,
the average emotional state of agents in that degree class is calculated
with more fidelity; more weight should be given to the corresponding degree class,
so that the value corresponding to the degree class is given more
importance in the approximation procedure.
We choose simple proportional weights which implies $w(d)=h(d)$, where
$h(d)$ represents the histogram of degrees (i.e., $h(d)$ is the number of
agents in the network with degree $d$).

The polynomials $p_i$ are thus chosen to be orthogonal
with respect to the degree distribution $h(d)=w(d)$.
This orthogonality condition can be described as follows:

\begin{equation}
<p_i(d),p_j(d)>_{w(d)} = \delta_{ij}.
\label{eq:Ortho}
\end{equation}

Since the polynomials are orthogonal with respect to the weight distribution,
the coefficients that minimize the residual (with respect to the same weights) in
Eq.~\ref{eq:MR} can be directly evaluated by the following simple expression:

\begin{equation}
c_i=<p_i(d),f(d)>_{w(d)}.
\label{eq:c}
\end{equation}
\begin{figure*}
\begin{center}
\resizebox{0.81\figwidth}{!}{%
  \includegraphics{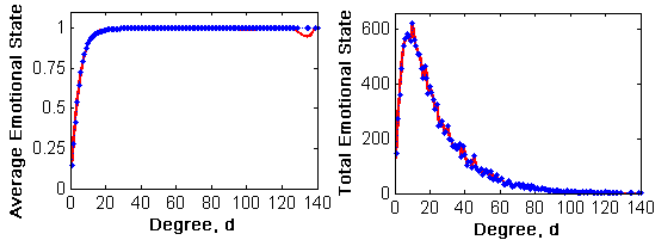}
}
\caption{
\textbf{Left:} The steady state average emotional state of all the
agents with a specified degree (as in the right plot of Fig.~\ref{fig:Ecorr})
is plotted versus the degree as (blue) dots.
The curve fit obtained by using $10$ of our orthogonal polynomials
is plotted as a (red) solid curve for comparison.
\textbf{Right:} For the same case, the {\em total} emotional state
of all the agents with a specified degree is plotted versus the
degree as (blue) dots.
This plot on the right is just a product of the plot on the left and
the degree histogram shown in Fig.~\ref{fig:Eav}.
The (red) solid lines correspond to the results from
our polynomial approximation procedure.
}
\label{fig:Fit}
\end{center}
\end{figure*}
What remains to be discussed is the selection of suitable polynomials.
For this portion, we borrow tools developed in the context of uncertainty quantification
in dynamical systems with random parameters.
In such problems, the effect of a random parameter (e.g. normally distributed) on the system state can be
represented through orthogonal polynomial expansions in terms of appropriate random variables \cite{Ghan91stochastic}.
These orthogonal polynomials are termed ``polynomial chaos" (PC), and projections
of the system states onto the Polynomial Chaos (PC) coefficients evolve deterministically
in time.
Such polynomials were initially proposed for Gaussian random parameters, but
a generalized approach for other random parameter probability measures
based on the Askey scheme was developed in \cite{Xiu02wiener--askey}.
These generalized polynomials, the ``Wiener-Askey polynomial chaos", are constructed
so that they are orthogonal with respect to a variety of known probability distributions.
In our case, the random parameter of interest {\em is the degree distribution itself}, and
it is a {\em truncated} geometric distribution.
If we approximate this as a (continuous) geometric distribution, the Wiener-Askey scheme suggests Meixner
polynomials as basis functions.
The derivation of orthogonal polynomials for any (discrete, possibly truncated/empirical) weight
function $w(d)$ (which we used for our numerical computations) is discussed in the Appendix.

In Fig.~\ref{fig:Fit}, we re-plot the curve of average emotional state versus degree
at the ``top" steady state branch (steady state $1$ in Fig.~\ref{fig:Eav})
using (blue) dots.
We evaluate the first $10$ polynomials that are orthogonal to the degree distribution
of our particular network, sampled empirically from the ``theoretical" truncated geometric distribution.
The corresponding $10$ coefficients are evaluated using Eq.~\ref{eq:c}
and our approximation of the system state $f(d)$ is then reconstructed using Eq.~\ref{eq:B} with $k=10$.
This reconstructed curve is plotted as a (red) continuous line in the same figure,
for comparison to the original curve.
Notice that the fit is more accurate at lower degree values, compared
to that at higher values; this is due to the small probability of high degrees - in other words,
due to the weight function used in our approximation procedure.
As can be seen in the left plot of Fig~\ref{fig:Eav}, the degree histogram is
heavily concentrated at lower degrees, and hence the states corresponding to these
degrees are better approximated in the reduction/reconstruction.
When we plot the {\em total} emotional state of a given degree class (the sum of the emotional states
of all the agents with the same degree) versus the degree, as shown on the right side of
Fig.~\ref{fig:Fit}, the error associated with our reduction/approximation procedure is clearly
much more uniformly spread across the degree classes.

\section{Coarse modeling and computational results}
\label{sec:ef}

Once the coarse variables of the model have been defined, the Equation-Free
approach \cite{Kevr04Equation-free:,Kevr03Equation-free} can be used
to computationally implement the coarse model.
The Equation-Free (EF) approach to modeling/computation is a framework
that has been developed for problems that can, in principle,
be described at multiple -here, two- levels.
The evolution equations
are available at a ``fine", microscopic scale (here at the level of
individual nodes/agents and edges/connections)
while the equations for the ``coarse", macroscopic behavior (here, a
handful of expansion coefficients for the function $f(d)$) are not available in an explicit, closed form.
In this approach, short bursts of simulations at the ``microscopic" node level are used to
estimate information such as time-derivatives, or actions of Jacobians,
pertaining to the coarse variables.
This is accomplished through the definition of operators that allow us
to ``translate" between coarse variables and consistent detailed, fine variables.
The transformation from coarse to fine variables is called
the {\em lifting} operator ($L$), while the reverse transformation
is called the {\em restriction} operator ($R$).

In our example, the fine variables of the model are the individual states
of all the agents in the social network, while the coarse variables are the
coefficients $c_i$ referred to in Sec.~\ref{sec:cr}.
The microscopic rules of evolution defined in Sec.~\ref{sec:model} constitute
the detailed, microscopic time-evolution operator defined by $\phi_t$
(where $t$ represents the number of time steps or iterations).
The macroscopic evolution operator can then be defined in terms of this  microscopic
evolution operator as well as the lifting and restriction operators as follows:

\begin{equation}
\Phi_t(\cdot) = R \circ \phi_t \circ L(\cdot).
\label{eq:Phi}
\end{equation}
In other words, the evolution of the coarse variables can be represented as:

\[ \mathbf{c}(T+t) = \Phi_t(\mathbf{c}(T)), \]
where $\Phi_t$ is defined in Eq.~\ref{eq:Phi} and
$\mathbf{c}$ represents the vector of coefficients [$c_1$, $c_2$,... $c_k$].

\begin{figure*}
\begin{center}
\resizebox{0.81\figwidth}{!}{%
  \includegraphics{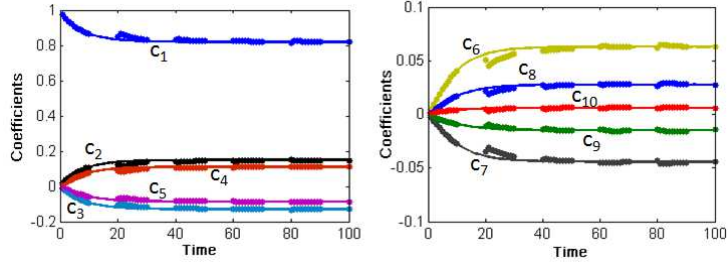}
}
\caption{
Evolution of the coarse variables corresponding to Fig.~\ref{fig:Ecorr}.
The solid lines indicate results obtained from direct simulations.
The dots indicate results obtained through coarse projective integration
using $10$ coefficients (accelerating the overall simulation by a factor of 2)
The plot on the left shows the evolution of the first $5$ coefficients,
while the plot on the right shows that of the next $5$ coefficients.
}
\label{fig:CPI}
\end{center}
\end{figure*}

We will illustrate the EF approach by describing two particular
algorithms: {\em coarse projective integration} \cite{Gear04computing,Rico04coarse}
and coarse fixed point computation \cite{Gear02`coarse'}.
In coarse projective integration, the detailed system is integrated forward in time using
short bursts of fine-scale, microscopic simulations, and the results are used to
estimate time derivatives of the coarse variables,
which are then used to project these latter variables forward in time.
Starting from a specified initial condition in terms of the ``microscopic"
variables (the emotional states of the agents), the following steps are
carried out to implement coarse projective integration:

\begin{enumerate}
\item \textbf{Simulation}: The detailed model is run for a specified period of time,
say $10$ time steps (to be exact, 1000 sample realizations are run and averaged).

\item \textbf{Restriction}: The coarse variables are evaluated at each time-step
of the simulation.
For our example, this involves computing the coefficients
$c_i$ given in Eq.~\ref{eq:c}.
\item \textbf{Projection}: The last few observations of the coarse variables $c_i$ from
the previous step are used to estimate their time-derivative.
The coefficients are then projected forward in time over another $10$ time steps
using a simple linear extrapolation.
Consider a standard differential equation $dx/dt = f(x)$.
The forward Euler scheme would read
           \[  x(t+\Delta t) \approx  x(t) + \Delta t f(x(t)). \]
Our projection is similar to this, except the time derivative ($f(x(t))$ above) does not come
from a closed formula ($f(x)$), but is instead estimated from the computational observations
of the fine scale model.

\item \textbf{Lifting}: Since estimates of the coarse variables $c_i$ are available at the
projected time, we have to transform them to a consistent set of agent states
for all the agents before the fine scale model can evolve again.
This can be achieved by expanding the polynomials according to Eq.~\ref{eq:B}.
Once the states of all the agents in the network are computed,
they can be used as the initial condition to restart simulations
as given in Step $1$.
Note that, if any of the agent states lie outside the acceptable range of values
(above $1$ or below $-1$), the states are revised to the limiting values
($1$ and $-1$ respectively).

\end{enumerate}
These steps, performed repeatedly, constitute coarse projective integration,
which is used to accelerate computations in systems with separations of time scales.
In our illustrative EF computations, we retained
all model parameters values mentioned above; for the coarse representation
we used $10$ polynomials, constructed to be orthogonal to
our network's degree distribution.
Using the corresponding $10$ coefficients as coarse variables,
we implemented coarse projective integration
for the trajectory shown in Fig.~\ref{fig:Ecorr}.
The evolution of the $10$ coefficients are shown as solid lines in
Fig.~\ref{fig:CPI}.
The first $5$ coefficients are shown on the left and the
next $5$ are shown on the right.
The results of coarse projective integration,
where fine scale simulations are carried out for only half of the computed trajectory
(leading to a factor of two in acceleration) are shown as dots for comparison in the same figure.
It can be seen that the coarse evolution can be captured well by our implementation of the
reduced model.

In addition to accelerating simulations,
we can use the Equation-Free framework to also quickly find coarse steady states
(corresponding to stationary states of the detailed, stochastically evolving model).
This is done using a coarse fixed point solver:
to compute the steady states values of the coefficients $\mathbf{c}$,
we have to solve the following equation:

\begin{equation}
F(\mathbf{c}) := \mathbf{c} - \Phi_{10}(\mathbf{c}) = 0,
\label{eq:NR1}
\end{equation}

where $\Phi_t$ is the coarse time-stepper over $t$ time steps.
The roots of $F$ can be found using a damped Newton-Krylov GMRES iteration scheme
\cite{Saad86gmres:,Kell95iterative}.
In the standard Newton-Raphson algorithm, the value of $\mathbf{c}$ would
be updated by the equation

\begin{equation}
\mathbf{c}_{n+1} = \mathbf{c}_n + \Delta \mathbf{c},
\label{eq:NR2}
\end{equation}

where $\Delta \mathbf{c}$ is found as the solution to

\begin{equation}
[DF(\mathbf{c}_n)]\Delta\mathbf{c} = -F(\mathbf{c}_n).
\label{eq:NR3}
\end{equation}

The action of the Jacobian $DF(\mathbf{c}_n)$ can be estimated indirectly
(since an analytical expression would require analytically differentiating
the coarse time-stepper $\Phi_{t}(\mathbf{c})$ in Eq.~\ref{eq:NR1}) using
the coarse time stepper.
We use a Krylov-based approach where the action
of this Jacobian on known vectors, its matrix-vector products, is
required to solve the equation;
this Jacobian can be estimated through numerical differentiation (i.e., evaluating the
coarse time-stepper at appropriately selected perturbations of the $\mathbf{c}$ values).
Such iterative {\em matrix-free} computations are naturally
suitable for Equation-Free computation, where explicit Jacobians are not
available in closed form.

\begin{figure*}
\begin{center}
\resizebox{0.9\figwidth}{!}{%
  \includegraphics{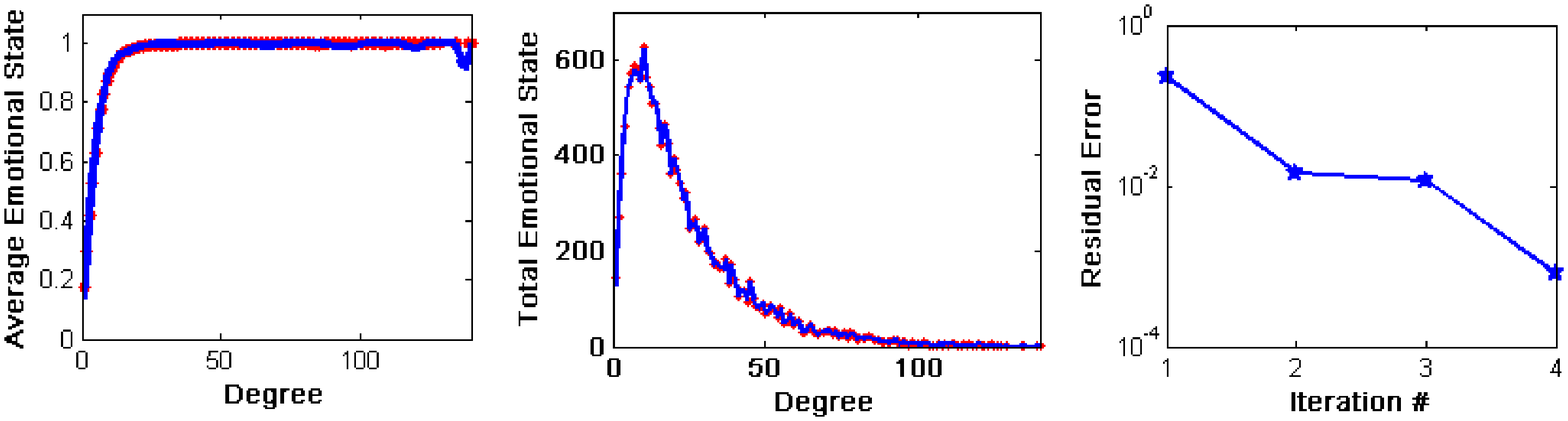}
}
\caption{
\textbf{Left:} The steady state
{\em average} emotional state of agents of a specified degree
versus the degree is plotted as (red) dots.
The lifting of the coarse steady state computed by
Newton-GMRES algorithm is plotted as a (blue) solid curve.
\textbf{Center:} The steady state
{\em total} emotional state of agents of a specified degree
versus the degree is plotted as (red) dots.
The lifting of the coarse steady state computed by
Newton-GMRES algorithm is plotted as a (blue) solid curve.
\textbf{Right:} Convergence of Newton-GMRES: The $L_2$ norm of
the coarse residual is plotted against the iteration number.
Computations were performed with a  $10$ polynomial
coarse basis.
}
\label{fig:NR}
\end{center}
\end{figure*}

The results of coarse fixed point computations are shown in Fig.~\ref{fig:NR}.
The first two plots in this figure show the actual steady state
(resp., the {\em average} and {\em total} emotional state of agents with a given degree
plotted against the degree class) computed using direct simulations as (red) dots.
The coarse steady state computed using the procedure described in the
previous section is plotted as a solid blue line.
The convergence of the norm of $F$ as defined in Eq.~\ref{eq:NR1} is
shown on the right plot of the same figure.

\begin{figure*}
\begin{center}
\resizebox{0.9\figwidth}{!}{%
  \includegraphics{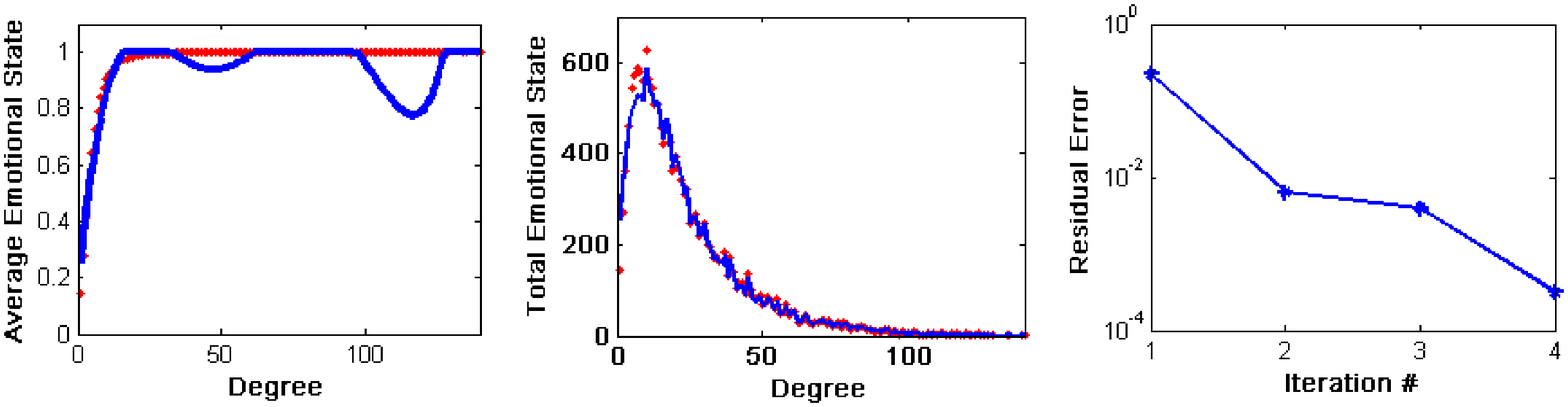}
}
\caption{
\textbf{Left:} The steady state
{\em average} emotional state of agents of a specified degree
versus the degree is plotted as (red) dots.
The lifting of the coarse steady state computed by
Newton-GMRES algorithm is plotted as a (blue) solid curve.
\textbf{Center:} The steady state
{\em total} emotional state of agents of a specified degree
versus the degree is plotted as (red) dots.
The lifting of the coarse steady state computed by
Newton-GMRES algorithm is plotted as a (blue) solid curve.
\textbf{Right:} Convergence of Newton-GMRES: The $L_2$ norm of
the coarse residual is plotted against the iteration number.
Computations were performed with a  $6$ polynomial
coarse basis.
}
\label{fig:NR6}
\end{center}
\end{figure*}

It is important to note here that although we need $10$ coefficients
to get a visually acceptable fit for the curve of average emotional state versus
degree, we can use an even smaller number of coefficients for an even more
parsimonious coarse-graining.
For instance, in Fig.~\ref{fig:NR6}, we show the results of coarse fixed point
computations through Newton-GMRES using just $6$ polynomials.
The plots in the figure are similar to the ones shown in Fig.~\ref{fig:NR}.
Although the polynomial fit truncated at $6$ coefficients
deviates visibly from the steady state function value,
especially at higher degrees (for reasons mentioned earlier),
(a) the plot on the right shows that the procedure
with $6$ coarse variables does converge,
and (b) the actual steady state can be easily recovered by
running simulations for a very short time
from the lifting of the computed coarse steady state.
Instead of truncating (setting higher order polynomial coefficients
in the reconstruction to zero), ``slaving" these coefficients to
the lower ones -in effect, constructing a six-dimensional ``slow manifold" for the
process dynamics- will provide even more accurate/more reduced models.

\section{Discussion}
\label{sec:conc}

In this work we discussed a systematic approach to coarse graining dynamics
of coupled networks of units (nodes, agents), and demonstrated its application
through the illustrative computational coarse-graining
of an agent-based model with an underlying network structure.
In this model, the state of the agents was observed to quickly become
highly correlated with the degree of the agents; our coarse graining methodology
takes advantage of this fact in constructing a set of appropriate
collective variables.
The resulting reduced model was computationally implemented
using the Equation-Free approach.

In the selection of our collective variables we took advantage of
concepts developed in the context of Uncertainty Quantification for studying
the effect of a random parameter in differential equations;
in effect, we are performing ``Heterogeneity Quantification" rather than
Uncertainty Quantification.
In UQ problems, the effect of a {\em random} parameter
with a known distribution on the system state is
captured by expanding the state in terms of orthogonal polynomials
of the random variables.
The orthogonal polynomials used depend on the distribution
from which the random variables are sampled.
In our case, the states of the agents depend on the agent structural identities
(here, agent degrees, whose distribution is prescribed by the network)
-and, of course, on time.
By analogy, we can think of the degrees as a ``random heterogeneity parameter"
with a given distribution, and parsimoniously capture its effect on the agent states
by expanding the states in terms of suitable orthogonal functions
of the degree.
It is clear that the approach can be extended to states that
depend on ``higher order" structural identities - identities
that do not only depend on the degree, but also on more/different
network statistics: for example, degree
{\em and} clustering coefficients for each node.
The joint distribution of these latter two features will again be dictated by the network,
and the basis functions will be now two-dimensional - clearly, the integrals
involved in computing the corresponding coefficients will start becoming
cumbersome as the number of ``determining features" grows.
For such problems, there has already been considerable progress on
collocation-based computations, and the use of sparse grids in the
UQ literature \cite{Xiu03modeling,Xiu09fast,Carl12managing} and we expect that these tools will
also become useful in network coarse-graining when multiple network features
affect the system state.
Still, there is no reason for the roots of polynomials orthogonal with
a given degree distribution weight to be themselves integers, and so
collocation approaches to approximating integrals over degree distributions
must be addressed.

We demonstrated how to computationally take advantage of
a coarse model through
coarse projective integration and coarse fixed point computations.
It is well known, and reasonably straightforward to extend the EF approach to other
system-level computational tasks, such as coarse bifurcation analysis, coarse
stability analysis etc. as reviewed in  \cite{Kevr03Equation-free}.
It is also worth mentioning that the approach described here is
in principle broadly applicable to coarse-graining other dynamical
problems consisting of heterogeneous coupled units.
Here the
relevant heterogeneity was the number of connections for each node (its degree);
in other network problems the heterogeneity may involve a non-network feature
(an ``intrinsic feature)
of the node (e.g., age or fitness of a node), or even possibly joint
distributions of ``network (structural)" and ``non-network (intrinsic)" heterogeneity features.
We are currently working on such ``joint heterogeneity quantification" problems.

We conclude by mentioning an additional, possibly important, consideration.
In recent literature on coupled oscillators (an area where our approach can
also be exploited) mean-field theories have been developed,
expressing the behavior of the network (the statistics of the distribution of
oscillator angles) as a function of the network degree \cite{Restrepo}.
What is of particular interest in our discussion is that, in order to derive
explicit mean field equations for behavior as a function of degree, the authors
could also include additional structure in the form of a prescribed ``degree
assortativity", the probability that a node with degree $k_i$ is connected
with a node of degree $k_j$; they also considered the case of {\em no} assortativity.
In our case, we do not derive such equations explicitly, but we solve them
through our equation-free approach. 
All our computations above were performed with a fixed, static network,
with a particular, prescribed degree distribution; choosing that particular
network, also {\em de facto} selected all additional high order statistics
through the network construction (including a particular assortativity).
In that sense, our equations constitute a coarse-graining {\em of the particular 
network}.
%
%

%
It is also conceivable that one may want to construct (and average over)
several sample networks {\em with the same statistics}, here the same
degree distribution (see, e.g., the
discussion in \cite{Goun11generation}).
Then the equation-free
approach does not solve for a coarse-graining of a particular realization,
but rather for the {\em expected} behavior over realizations with a prescribed 
degree distribution.
Creating several such network samples will be used to estimate
the common degree distribution needed to compute our orthogonal polynomial basis.

It is thus our {\em lifting step} that determines what coarse-grained problem we are solving.
Lifting to always the same network (resp. lifting to networks with prescribed
degree distributions only (and averaging over them), resp. lifting to networks with
prescribed degrees and assortativities (and averaging over them)) yields
different coarse-grained problems. 
Our approach can be used to tackle all these problems (and obvious
variations/extensions of them) by judiciously selecting what subset of realizations is constructed
in the lifting step.

It is finally conceivable that the approach may be
extended to encompass cases where not only the state of the nodes, but the
statistics of the network itself may evolve over time - so that not only
dynamics {\em on} networks -like here-, but {\em adaptive network} dynamics
(dynamics simultaneously {\em on} and {\em of} networks) could be tackled.

{\bf Acknowledgements} This work was partially supported by the US AFOSR and
by the US Department of Energy. I.G.K and C.R.L. grateful to the Institute for Advanced Study, T. U. Muenchen for support by a Hans Fischer Senior Fellowship to I.G.K.; C.I.S acknowledges support by a Fulbright Fellowship for a research visit to Princeton.

\appendix

\section{Finding a suitable basis of orthogonal polynomials tailored to a given degree distribution}
\label{sec:app}

The procedure that we use to evaluate a set of polynomials orthogonal to
one another with respect to an empirical weight distribution \cite{Pres92numerical}
(defined over a range of integers, here the node degrees) is described here.
Let the $i$-th required polynomial be denoted by $p_i$, and let $w(d)$ be the specified
discrete weight distribution.
$p_i$ can be written using the following general representation:

\begin{equation}
p_i=\alpha_i \left( 1+\sum_{j=1}^{i-1}y_{ij}d\right).
\label{eq:A1}
\end{equation}
The orthogonality condition is written as

\begin{equation}
<p_i,p_j>_{w(d)} = \delta_{ij}.
\label{eq:A2}
\end{equation}
Since we are interested in evaluating the function
at discrete values of $d$, we may approximate
the orthogonality condition by the following summation:

\begin{equation}
\sum_{d=1}^{140} p_i(d)p_j(d)w(d) = \delta_{ij}.
\label{eq:A3}
\end{equation}
We are interested in finding the first $k$ polynomials.
This implies that we need to evaluate $k(k+1)/2$ polynomial coefficients,
i.e., $k(k-1)/2$ for $y_{ij}$s and $k$ for $\alpha_i$s.
These can be {\em successively} evaluated by using the $k(k+1)/2$ orthogonality conditions:
$k(k-1)/2$ for the case $i\neq j$ and $k$ for the case $i = j$ respectively.
%


\end{document}